# Distribution of Electron Charge Centres:

# A Picture of Bonding Based on Geometric Phases

Joydeep Bhattacharjee, Shobhana Narasimhan and Umesh V Waghmare

*Theoretical Sciences Unit,*

*Jawaharlal Nehru Centre for Advanced Scientific Research,*

*Jakkur PO, Bangalore 560 064, India*

In the past two decades, geometric phases[1-3] have provided a powerful new way of looking at quantum mechanical systems, manifesting themselves in subtle but observable ways[4-6]. Here, we use them to define a versatile function ("distribution of electron charge centres" or DECC) which can be easily evaluated and interpreted, providing information about electronic structure in real space. Its utility is illustrated by application to a large variety of insulators, metals and molecules, treated here within the framework of density functional theory. The DECC is shown to provide a precise and compact description of bonding. Unshared-electron (ionic) and shared-electron (covalent, metallic) bonds are shown to present clearly distinct signatures: the former are uni-centred while the latter are $n$-centred ($n \geq 2$). Moreover, the charge contained in the DECC peaks gives either the ionic charge or the number of shared electrons, which is an even integer for covalent bonds. One obtains revealing insight into the microscopic chemical origins of macroscopic phenomena such as ferroelectricity[7] in $PbTi0_3$ and the anomalous mechanical behaviour[8] of bulk Al relative to that of Cu.

Our work represents a new step in a tradition of attempts to characterize and classify chemical bonds, e.g., Lewis' valence theory[9], localized molecular orbitals[10-13], Bader's topological analysis of charge densities[14], and Silvi and Savin's introduction[15] of the electron localization function or ELF. Looking at information about the mean position of electrons, derived rigorously from geometric phases, provides a radically different approach to the study of bonding. Compared to previous tools, the DECC provides a wider variety of information, while also offering a more easily visualized and quantifiable picture of bonding. The existence of sharply localized features in the DECC function makes the characterization of bonding, charge transfer, etc., a well-defined and easy procedure, as ambiguities regarding volumes of integration do not arise, nor does one have to search for critical points in the charge density. Another recent technique that leads to a visual picture of bonding is the computation of maximally localized Wannier functions[16,17], introduced by Marzari and Vanderbilt[18]. With no specific choice of Wannier functions, the DECC provides similar information, but in a more compact way, as one does not need to perform a band-by-band analysis; it is also more readily obtained as it avoids a variational calculation.

When a Bloch electron in a periodic system evolves through a full period in reciprocal space, it acquires a memory of its average position, in the form of a geometric phase[4-6]. This has formed the basis of methods to calculate the electric polarization related properties of insulating crystals from first principles[5,6,18]. We now aim to extend these advances to obtain a simple, concise and precise description of bonding in real space.

One would like to obtain maximally localized orbitals by diagonalizing the position operator $\mathbf{r} = (x,y,z)$ in the subspace of occupied electronic levels[19,20]. Eigenvalues of the operator $Pr_iP$, where the operator $P$ projects onto the occupied subspace, can be obtained for a periodic system through computation of eigenvalues $\tau_i^i(\mathbf{k})$ of a geometric phase matrix $\Gamma_i(\mathbf{k})$ (see Supplementary Online Material (SOM)). While this can be pursued in one dimension, this procedure cannot formally be extended to three dimensions, since though the position operators $x$, $y$ and $z$ commute, the three operators $(Pr_1P, Pr_2P, Pr_3P)$ or $(\Gamma_1(\mathbf{k}), \Gamma_2(\mathbf{k}), \Gamma_3(\mathbf{k}))$ do not, in general, commute and therefore can not be diagonalized simultaneously[18]. This leads to approximations when evaluating maximally localized Wannier functions in three dimensions We avoid this problem by making use of quantum mechanical joint distribution functions[21] for noncommuting operators and define a "distribution of electron charge centres" (DECC):

$$D(\mathbf{r}) = \int d\mathbf{k} \sum_{lmn} \delta(\mathbf{r} - \mathbf{T}_{lmn}(\mathbf{k})) \left\langle v_{\mathbf{k}l}^1 \middle| v_{\mathbf{k}m}^2 \right\rangle \left\langle v_{\mathbf{k}m}^2 \middle| v_{\mathbf{k}n}^3 \right\rangle \left\langle v_{\mathbf{k}n}^3 \middle| v_{\mathbf{k}l}^1 \right\rangle , \qquad (1)$$

where the integral runs over the first Brillouin zone (BZ), and the summation is over all occupied electronic levels. $v_{\mathbf{k}m}^i$ and $\tau_m^i(\mathbf{k})$ are the eigenfunctions and eigenvalues respectively of the geometric phase matrix $\Gamma_i(\mathbf{k})$, where $i$ runs over the three Cartesian directions, and $\mathbf{T}_{lmn}(\mathbf{k}) = (\tau_l^1(\mathbf{k}), \tau_m^2(\mathbf{k}), \tau_n^3(\mathbf{k}))$. It is fairly straightforward to generalize this definition to the case of metals, by including occupation factors; see the SOM. Note also that though such joint probability functions need not be positive definite[21], $\rho(\mathbf{r})$ is still correctly normalized: $\int_C d\mathbf{r} \rho(\mathbf{r}) = N_e$, where $C$ represents the periodic unit cell of volume $\Omega$, containing $N_e$ electrons. Moreover, the first moment of $D$ gives the electronic polarization $\mathbf{P}_e = (1/\Omega) \int_C d\mathbf{r} D(\mathbf{r}) \mathbf{r}$.

Evaluating the DECC for a system involves using a simple post-processing routine that can be appended to any of the standard *ab initio* density functional theory (DFT) packages. $D(\mathbf{r})$ exhibits peaks at the average positions of electrons, giving information about the location of electrons, similar to that provided by crystal structure about the location of nuclei.

Upon computing $D(\mathbf{r})$ for systems consisting of a single atom, we find that it displays a clear shell structure, i.e., peaks corresponding to orbitals with different principal quantum numbers fall in different spatial shells. For molecules and crystals, the spatial connectivity of electronic and atomic average positions allows for a natural classification and quantitative characterization of bonding. Typically, there is a cluster of peaks associated with the centre of each bonding orbital or a Wannier function. The width of these peaks is a measure of the uncertainty in determining all three coordinates of the average position of a Bloch electron simultaneously.

As a first example of the utility of the DECC, we consider the case of ionic bonding (Figure 1). For such systems, the DECC exhibits a set of peaks surrounding each ion, with its local point symmetry. An ionic bond shows up as a *uni-centred* bond in the present approach. The static charge of an ion, obtained unambiguously by summing up the charge contained in these sharply localized peaks, is in general found to be *identical* to the expected nominal charge. For example, for paraelectric $PbTiO_3$ (Fig 1c), one obtains charges of +2, +4 and -2 at the sites of Pb, Ti and O, respectively. Dynamical charges are obtained by evaluating the first moment of $\Delta D(\mathbf{r})$, the *change* in the DECC arising from a small atomic displacement. Evaluating $\Delta D(\mathbf{r})$ as one transforms from the paraelectric to ferroelectric structure of $PbTiO_3$ , via an off-centring displacement of Ti atoms (Figures 1c and 1d), yields an accurate value of 7.1 for the anomalous Born dynamical charge of Ti. Moreover, one can identify its origin as being a shift in the centre of O *p*-like orbitals perpendicular to the Ti-O bonds. There are no new peaks at the O-site, so this shift cannot be due to local polarizability, but must arise from a redistribution of the small charge in the Ti *d*-states contributing to these bonding orbitals, which has been shown to be the origin of ferroelectricity[7].

For covalently bonded systems, $D(\mathbf{r})$ is usually found to be *two-centred,* i.e., it has peaks on the lines connecting pairs of atoms (see Figure 2). For non-polar covalent bonds, these peaks are at the centre of the bond, whereas for polar bonds, the peaks are shifted towards the atom that is more electronegative. Interestingly, we find that *only* for covalent systems, $D(\mathbf{r})$ takes on a sizeable negative value at some points; this corresponds to a violation of the D'Espagnat inequality[21, 22] in quantum mechanics, indicating that covalent bonds can arise only in a quantum mechanical description.

In contrast, metallic bonding is found to exhibit a clear *multi-centred* character. Peaks in $D(\mathbf{r})$ containing *m* electrons and enclosed in an *n*-atom polyhedron, define *n*-centred *m*-electron bonds (see Figure 3). For some metals, there are also additional peaks at atomic sites. For Mo, we find a charge of 9.02 *e* at each of the two atomic sites, and 1.66 *e* at the centre of each of the six octahedral holes in the unit cell, where *e* is the unit of electronic charge. The latter correspond to the second-neighbour bonds[23] in Mo, relevant to the brittleness of BCC metals at low temperatures. The positioning of the peaks is sensitive to both crystal structure and electronic structure. For example, Al, Cu and Pb are all face-centred cubic (fcc) metals, but their DECCs exhibit differing features. For Al, we find the most prominent feature (with a charge content of 1.40 *e*) is at the centre of each $Al_4$ tetrahedron; there are secondary features (with a charge content of 0.04 *e*) halfway between nearest-neighbour atoms. In contrast, for Cu and Pb, most of the charge (10.4 *e* for Cu and 12.58 *e* for Pb) is contained in peaks at each of the atomic sites, with secondary peaks (containing 0.6 *e* for Cu and 1.42 *e* for Pb) at the centre of each octahedral hole. This difference testifies to the existence of more covalent (i.e., greater number of shared electrons per atom) bonding in Al than in Cu. This has been invoked as an explanation[8] of a puzzling anomaly: Al has a higher mechanical shear strength than Cu, even though the shear modulus of Al is lower than that of Cu. To further investigate this issue, we examine the change in the bonding in Al and Cu at an intrinsic stacking fault in (111) planes (see Figure 3e), which is relevant to the shear deformation and strength of fcc metals. We find that the covalent bonds in Al are significantly altered at the stacking fault, as shown by a fragmentation of the DECC peaks at $Al_4$ centres. Such a fragmentation implies the involvement of excited states in bonding. In contrast, for Cu, the already weak DECC peaks at $Cu_6$ centres are only slightly affected at the stacking fault. Correspondingly, the energy cost associated with formation of a stacking fault is much higher for Al than Cu, resulting in the higher mechanical shear strength of the former.

Finally, we show that the DECC captures bonding in molecules, by presenting a few examples (Figure 4); these results follow the same trends as displayed for extended solids. For example, the covalent C-C σ-bonds in organic molecules are manifested as peaks in $D(\mathbf{r})$ at the bond-centres, whereas for C-H bonds the peaks are closer to the H atoms. The C-C π-bonds in the case of $C_2H_4$ and $C_2H_2$ show up as peaks that are symmetrically off-centred in the perpendicular bisecting planes of the C-C bonds. Note that double and triple bonds look clearly different, and the estimate of the bond order for the C-C bond in the case of $C_2H_6$, $C_2H_4$ and $C_2H_2$ is 1, 2 and 3 respectively. The three-centre bond in $B_2H_6$ manifests itself in a DECC peak located inside the B-H-B triangle (closer to the H atom).

*Methods:* The results presented above were all obtained by performing standard plane-wave pseudopotential *ab initio* density functional theory calculations using the CASTEP 2.1[24] and ABINIT[25] packages. In order to evaluate the geometric phase matrices Γ, one needs parallel-transported Bloch wave functions[20], which have been obtained using discretized parallel transport based on energy eigenfunctions[18]. $D(\mathbf{r})$ was calculated on a real-space mesh and was visualized using XCrysDen software[26]. For graphical purposes, we have plotted isosurfaces that are typically at 10% of the maximum value.

To summarize: we have defined a function derived from the geometric phases of electronic wavefunctions, which provides information in an easily evaluable and interpretable way, about the distribution of electrons in real space. It provides a clear visual picture and classification of bonding, and makes possible the unambiguous determination of both static and dynamical charges. In this letter, we have confined ourselves to familiar systems, with the goal of presenting evidence for the reliability and reasonableness of DECC-derived information. We foresee that the DECC can be used as a simple but powerful tool to understand and design novel functional materials.

References:

1. Berry, M. V., *Proc. R. Soc. London Ser, A* **329**, 45-57(1994).

2. Anandan, J., *Nature* **360**, 307-313 (1992).

3. Shapere, A. & Wilczek, F., *Geometric Phases in Physics* (World Scientific, 1989).

4. Zak, J., *Phys. Rev. Lett.* **62**, 2747-2750 (1989).

5. King-Smith, R. D. & Vanderbilt, D., *Phys. Rev. B* **47**, 1651-1654 (1992).

6. Resta, R., *Rev. Mod. Phys.* **66**, 899-915 (1994).

7. Cohen, R. E. *Nature* **358**, 136-138 (1992).

8. Ogata, S., Li, J. & Yip, S., *Science* **298**, 807-811 (2002).

9. Lewis, G. N., *Valence and the Structure of Atoms and Molecules* (Dover, New York, 1966).

10. Foster, J. M. & Boys, S. F., *Rev. Mod. Phys.* **32**, 300-302 (1960).

11. Ruedenberg, K., *Rev. Mod. Phys.* **34**, 326-376 (1962).

12. Edmiston, C. & Ruedenberg, K., *Rev. Mod. Phys.* **35**, 457-464 (1963).

13. Pauling, L,. *The Nature of the Chemical Bond* (Cornell Univ. Press, 1960).

14. Bader, R. F. W. *Atoms in Molecules: A Quantum Theory* (Oxford University Press, 1990).

15. Silvi, B. & Savin, A., *Nature* **371**, 683-686 (1994).

16. Wannier, G., *Phys. Rev.* **52**, 191-197 (1937).

17. Kohn, W.*, Phys. Rev. B* **7**, 4388-4398 (1973).

18. Marzari, N. & Vanderbilt, D. *Phys. Rev. B* **56**, 12847-12865 (1997).

19. Sgiarovello, C., Peressi, M. & Resta, R. *Phys Rev. B* **64**,115202*(10pages)* (2001).

20. Bhattacharjee, J. & Waghmare, U. V. *Phys. Rev. B* **71**, 045106*(5pages)* (2005).

21. Barut, A. O., Bozic, M. & Maric, Z. *Foundations of Physics*, **18**, 999-1020 (1988).

22. D'Espagnat, B. *Sci. Am.* **241**, 158-167 (1979).

23. Eberhart, M. E. *Prog. in Surface Science.* **36**, 1-34 (1991)

24. Payne, M. C. *et al.* CASTEP 2.1, Cavendish Laboratory, University of Cambridge.

25. First-principles computation of material properties : the ABINIT software project. Gonze, X., *etal. Computational Materials Science* **25***, 478-492 (2002).*

26. Kokalj, A. XCrySDen. *J. Mol. Graphics Modelling*, **17**, 176-179 (1999).

ACKNOWLEDGEMENT. We thank K. S. Narayan, Karin Rabe, M. S. Narasimhan and Mas Subramanian for a critical reading of the manuscript. JB thanks CSIR, India for a research fellowship.

UVW acknowledges support from the DuPont Young Faculty Award. Some of the calculations were carried out in the central computing facility of JNCASR, funded by the Department of Science and Technology, Government of India. We also thank the anonymous referee whose constructive criticism helped us in substantially improving the methodology to obtain DECC, as presented in this manuscript.

Figures :

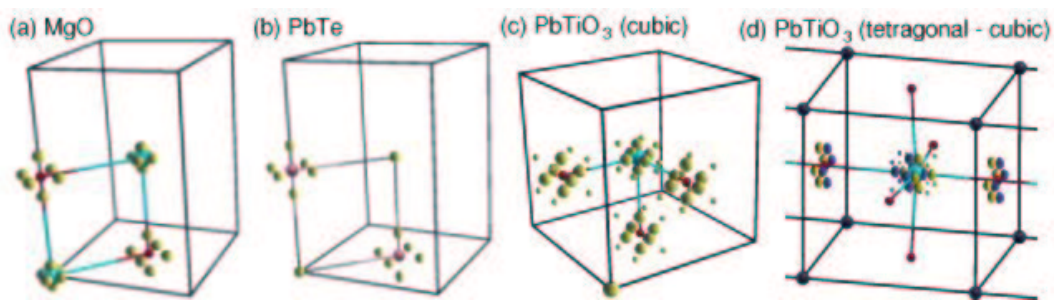

Figure 1: Isosurfaces of DECC for some ionic insulators: (a) Mg0 (b) PbTe (c)cubic (paraelectric) PbTiO$_3$. All unit cells have been drawn such that the cations are at the corners and centre of the cell. In all the cases, the DECC peaks are found to be centred at ionic sites, and can be assigned to specific electronic states. For MgO (a), similar peaks at Mg and O sites correspond to (2s 2p) semicore and valence states respectively, with the peaks at Mg closer to the nucleus than those at O. In isostructural PbTe (b), a similar feature is found near the Te site. The isolated peaks at the Pb sites in PbTe and paraelectric PbTiO$_3$ (c) correspond to the lone pair of 6s electrons. The peaks at the Ti and O sites in PbTi0$_3$ correspond to (3s, 3p) and (2s,2p) states respectively. In (d), we have plotted $\Delta D(\mathbf{r})$ when paraelectric PbTi0$_3$ is distorted to form ferroelectric PbTi0$_3$. Yellow and dark blue peaks denote positive and negative values respectively.

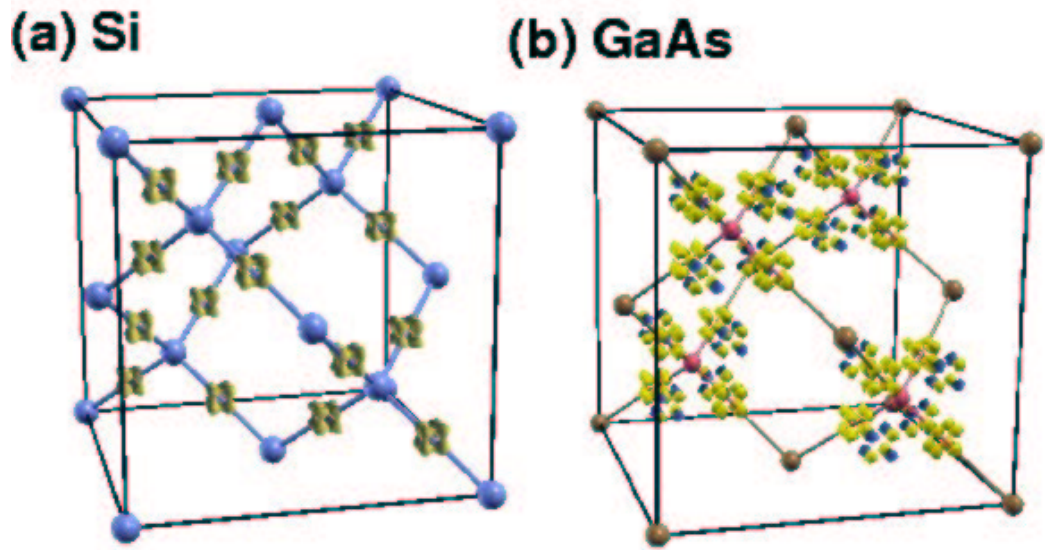

Figure 2: Isosurfaces of DECC for covalent crystals: (a) Si and (b) GaAs. Light blue, red and brown spheres denote Si, As and Ga atoms respectively. Yellow /dark blue peaks represent positive/negative values of DECC. There are large positive-valued peaks at the centre of the homopolar Si-Si bonds, whereas they are shifted closer to As for the heteropolar Ga-As bonds. The bond order obtained by summing up the charges in these peaks for each of these bonds is 1. In addition, at the antibonding sites in GaAs, there are both positive-valued and negative-valued peaks, the total charge contained in them being zero. (Similar peaks are present in Si too, but they are visible at much lower absolute isosurface values compared to those in GaAs.)

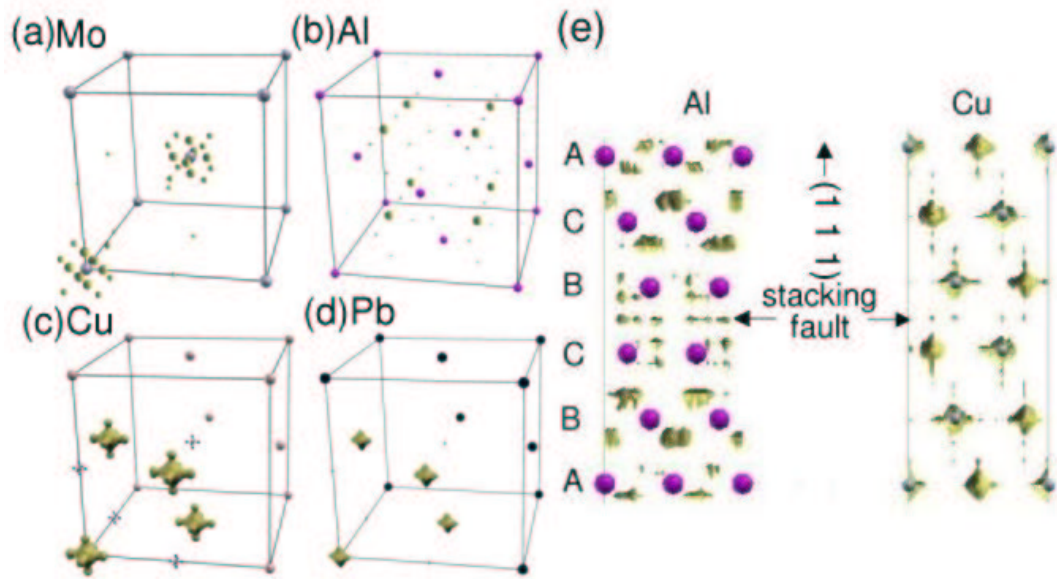

Figure 3: Isosurfaces of DECC for metals: (a) Mo (b) Al (c) Cu and (d) Pb. Though all of the latter three have FCC structure, the DECC peaks in them look quite different. Prominent peaks at the Mo atomic site correspond to $4s$ and $4p$ states and there are bonding peaks at the centres of second-neighbour bonds. In Al the dominant bonding peaks are at the tetrahedral holes with no peaks at the Al site. The peaks on the atoms in Cu and Pb correspond to mostly the $d$ electrons. Bonding peaks at the octahedral holes of Cu and Pb are much weaker compared to those in Al. In (e) we show the changes in the DECC structure upon introduction of stacking fault in the contrasting cases of Al and Cu.

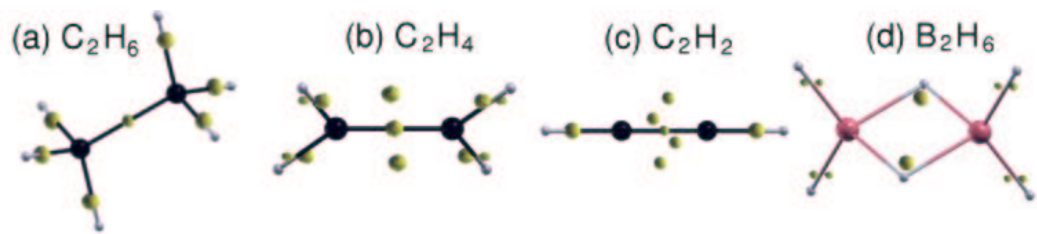

Figure 4: Isosurfaces of DECC for molecules, showing C-C single ($C_2H_6$), double ($C_2H_4$), and triple ($C_2H_2$) bonds, and the three-centre two-electron bond in $B_2H_6$.